\documentclass[aps,prx,preprint,amsmath,preprintnumbers,amssymb,aps,superscriptaddress]{revtex4-2}

\usepackage{graphicx}
\usepackage{dcolumn}
\usepackage{bm}
\usepackage{hyperref}
\hypersetup{colorlinks=true, linkcolor=blue, linktocpage}
\usepackage{amsmath, amssymb, 
,setspace}
\usepackage[pdftex]{color
}
\usepackage{epsfig}
\usepackage{xcolor}
\usepackage{physics}
\usepackage{amsfonts}
\usepackage{diagbox}

\begin{document}
%
%
%
\newcommand{\ion}[2]{\mbox{$^{#2}$#1$^+$}}
\newcommand{\Ca}[1]{\ion{Ca}{#1}}
\newcommand{\Be}[1]{\ion{Be}{#1}}
\newcommand{\Al}[1]{\ion{Al}{#1}}
\newcommand{\Hg}[1]{\ion{Hg}{#1}}
\newcommand{\Mg}[1]{\ion{Mg}{#1}}
\newcommand{\Cd}[1]{\ion{Cd}{#1}}
\newcommand{\Yb}[1]{\ion{Yb}{#1}}
\newcommand{\lev}[2]{\mbox{#1$_{\mbox{\tiny$#2$}}$}}
\newcommand{\hfslev}[3]{\mbox{#1$^{\mbox{\tiny$#3$}}_{\mbox{\tiny$#2$}}$}}
%
\newcommand{\unit}[1]{\,\mbox{#1}}
\newcommand{\mHz}{\unit{mHz}}
\newcommand{\Hz}{\unit{Hz}}
\newcommand{\kHz}{\unit{kHz}}
\newcommand{\MHz}{\unit{MHz}}
\newcommand{\GHz}{\unit{GHz}}
\newcommand{\THz}{\unit{THz}}
\newcommand{\torr}{\unit{torr}}
\newcommand{\mV}{\unit{mV}}
\newcommand{\mW}{\unit{mW}}
\newcommand{\uW}{\unit{$\mu$W}}
\newcommand{\mrad}{\unit{mrad}}
\newcommand{\cm}{\unit{cm}}
\newcommand{\mm}{\unit{mm}}
\newcommand{\um}{\unit{$\mu$m}}
\newcommand{\nm}{\unit{nm}}
\newcommand{\K}{\unit{K}}
\newcommand{\s}{\unit{s}}
\newcommand{\cps}{\unit{s$^{-1}$}}
\newcommand{\persec}{\unit{s$^{-1}$}}
\newcommand{\ms}{\unit{ms}}
\newcommand{\us}{\unit{$\mu$s}}
\newcommand{\ns}{\unit{ns}}
\newcommand{\uA}{\unit{$\mu$A}}
\newcommand{\degree}{\mbox{$^{\circ}$}}
\newcommand{\degC}{\mbox{\degree{}C}}
\newcommand{\G}{\unit{G}}
\newcommand{\mG}{\unit{mG}}
%
\newcommand{\citesec}[2]{\cite[\S{}#2]{#1}}   
\newcommand{\eg}{{\em e.g.}}
\newcommand{\ie}{{\em i.e.}}
\newcommand{\viz}{{\em viz.}}
\newcommand{\ish}{\mbox{$\sim$}\,}
\newcommand{\ltish}{\protect\raisebox{-0.4ex}{$\,\stackrel{<}{\scriptstyle\sim}\,$}}
\newcommand{\gtish}{\protect\raisebox{-0.4ex}{$\,\stackrel{>}{\scriptstyle\sim}\,$}}
\newcommand{\lr}{\mbox{$\leftrightarrow$}}
\newcommand{\diff}[1]{\mbox{\/d$#1$}}
\newcommand{\smalltriangle}{\protect\raisebox{0.2ex}{\tiny $\triangle$}}
\newcommand{\blob}{\mbox{$\bullet$}}
\newcommand{\file}[1]{{\small\tt #1}}
\newcommand{\wee}[2]{\mbox{$\frac{#1}{#2}$}}
\newcommand{\sub}[1]{\mbox{$_{\mbox{\tiny #1}}$}}
\newcommand{\intl}{\int\limits}
\newcommand{\vep}{\varepsilon}
\newcommand{\dagg}{^\dagger}
%
\newcommand{\qs}{\marginpar{\tt S}}          
\newcommand{\qr}{\marginpar{\fbox{\tt R}}}   
\newcommand{\ql}{\marginpar{\fbox{\tt !}}}   
\newcommand{\qn}[1]{$\rightarrow${\tt #1}}   
\newcommand{\qv}[1]{\\{}\fbox{\parbox{\textwidth}{\tt #1}}}   
\newcommand{\qnew}{\marginpar{\fbox{\tt NEW}}}
\newcommand{\xref}{\marginpar{\tt X}}        
\newcommand{\tyler}[1]{\textcolor{black!30!cyan}{\textbf{TG:} #1}}
\newcommand{\didi}[1]{\textcolor{black!30!green}{\textbf{DL:} #1}}
%
%
\preprint{LLNL-JRNL-859047}

\title{Fast Ground State to Ground State Separation of Small Ion Crystals}

\author{Tyler H. Guglielmo}
\email{guglielmo2@llnl.gov}
\affiliation{Lawrence Livermore National Laboratory, Livermore CA}
\author{Dietrich Leibfried}
\affiliation{National Institute of Standards and Technology, Boulder CO}%
\author{Stephen B. Libby}%
\affiliation{Lawrence Livermore National Laboratory, Livermore CA}%
\author{Daniel H. Slichter}
\affiliation{National Institute of Standards and Technology, Boulder CO}%

\date{\today}

\begin{abstract}
Rapid separation of linear crystals of trapped ions into different subsets is critical for realizing trapped ion quantum computing architectures where ions are rearranged in trap arrays to achieve all-to-all connectivity between qubits.  We introduce a general theoretical framework that can be used to describe the separation of same-species and mixed-species crystals into smaller subsets.  The framework relies on an efficient description of the evolution of Gaussian motional states under quadratic Hamiltonians that only requires a special solution of the classical equations of motion of the ions to describe their quantum evolution under the influence of a time-dependent applied potential and the ions' mutual Coulomb repulsion. We provide time-dependent applied potentials suitable for separation of a mixed species three-ion crystal on timescales similar to that of free expansion driven by Coulomb repulsion, with all modes along the crystal axis starting and ending close to their ground states. Three separately-confined mixed species ions can be combined into a crystal held in a single well without energy gain by time-reversal of this separation process.
\end{abstract}

\maketitle
\newpage


\section{Introduction}
As trapped ion quantum information processors continue to evolve and scale up, efficient all-to-all connectivity becomes increasingly valuable.  The quantum charge-coupled device architecture (QCCD)~\cite{wineland1998experimental} laid out a path for scaling ion trap computers beyond single chains to universal quantum information processors with all-to-all connectivity~\cite{wineland1998experimental,kielpinski2002}.  In recent cutting-edge experimental implementations of the QCCD architecture, the separation, transport, recombination, and re-cooling of ions consumes $98~\%$ or more of the runtime of representative algorithms~\cite{Pino_2021,moses2023race}.  Separation, transport, and recombination can be classified by whether they occur on timescales lasting many periods of the ion motional modes (adiabatic) or on timescales on the order of the motional period (faster-than-adiabatic or diabatic). Faster-than-adiabatic ion transport with motional modes starting and ending near the ground state has been demonstrated experimentally for crystals of one or two ions of the same species~\cite{Bowler2012Transport,walther2012,Sterk:2022ruj,clark2023,Kaufmann_2017}. However, separation of multi-ion, multi-species crystals into two subsets with motional modes starting and ending in the ground state (ground state to ground state separation, or GGS) has only been demonstrated in the adiabatic regime, which is inherently slow~\cite{rowe2002,barrett2004,home2009,Bowler2012Transport,Blakestad_2009,Wan_2019,Lancellotti:2023bgn,Pino_2021,moses2023race}.  Theoretical work exploring fast GGS has so far been limited to single species or single ion crystals \cite{Torrontegui_2011,Lau_2011,Kaufmann_2014,Palmero_2014,Palmero_2015,Ruster_2014,Lau_2014,Lizuain_2017,Simsek:2021dkp}.  Furthermore, to achieve many key metrics required of efficient quantum computers, one must be able to achieve efficient GGS.  For example, high gate fidelities and short recooling times are difficult or impossible without GGS.  Hence, GGS is necessary for essentially all ion connectivity reconfigurations in the QCCD architecture.  As such, finding and implementing faster-than-adiabatic GGS protocols is crucial for removing a major speed limitation in current ion trap quantum information processors.\\
\\
A method for fast GGS was introduced in Ref.~\cite{Sutherland:2021ifb}, wherein two same species ions initially in a single potential well of a linear rf ion trap were allowed to fly apart through their mutual Coulomb repulsion by rapid removal of the applied confining potential, and then were re-trapped in diabatically-applied separate potential wells. GGS was achieved in this scenario through special initial state preparation of the center-of-mass (COM) and stretch normal modes along the crystal axis (axial modes).  In particular, the states were squeezed prior to the release of the ions in the initial trap.  Since the ion motional states evolve under approximately quadratic Hamiltonians during separation, the unitary time evolution of each mode, from release to recapture, can be represented by a combination of one squeeze operator and one rotation operator, i.e. the Euler decomposition of $\text{SU}(1,1) \cong \text{SP}(2,\mathbb{R})$ \cite{Ehrenfest_1927,Heller_1975,Huber_1987,BLOCH196295,SudarshanSimon,Perelomov_1972,Wodkiewicz:85,Gerry_1985,Yurke_1986,Wu_2006}.  This decomposition allows one to prepare each mode in advance of separation with a squeezing operation that will be exactly undone by the effects of separation, thus enabling capture of the ions in their axial motional ground states after diabatic separation.\\
\\
Our work extends Ref.~\cite{Sutherland:2021ifb} to the case of more general linear ion crystal configurations that may contain more than two ions and more than one species. This is accomplished through a combination of squeezing and beam-splitting, i.e. the Bloch-Messiah \cite{BLOCH196295} decomposition of $\text{SP}(4,\mathbb{R})$ \cite{Weedbrook_2012,SudarshanSimon,SIMON1987223,Arvind_1995}. In addition, rather than pre-squeezing the axial modes of an ion crystal before separation, we identify suitable time-dependent applied potentials that continuously transform all modes during separation such that they end up close to their ground states in their final potential wells. This ``on the fly'' transformation should result in further reduction of GGS durations and manifests a method for rapid squeezing that is different from squeezing protocols previously implemented for ion motional modes~\cite{heinzen1990,meekhof1996,kienzler2015,Burd_2019}. We present a practically important example, the GGS of a same-species or mixed-species data-helper-data (DHD) three-ion crystal into three separate wells, and lay out the principles for GGS of larger same- or mixed-species crystals with additional axial modes.  The method is general to any choice of data and helper qubit ions, but we specialize to the case of Be$^+$-Mg$^+$-Be$^+$ (BMB) for concreteness when discussing a practically relevant example.\\
\\
In Section~\ref{sec:Formalism} we introduce a formalism that is convenient for analyzing Gaussian states and their evolution under quadratic Hamiltonians and will be used throughout the paper.  Section~\ref{sec:TransProt} applies this formalism to characterize the squeezing and beam-splitting operations acting on the axial normal modes of a DHD crystal during a swift ramp down of the external potential to zero as well as an equally swift ramp-up of three separate ``catching wells''\textemdash arresting the three separated ions in analogy to the protocol discussed in Ref.~\cite{Sutherland:2021ifb}. We then outline the approach for even larger crystals in \ref{sec:LargeCrystals}. Previously, all state preparation had been assumed to have occurred by an unknown mechanism \emph{prior} to separation \cite{Sutherland:2021ifb} or that the modes are returned to their ground states \emph{after} separation is complete.  In Sec.~\ref{sec:fastThreeIon} we overcome this limitation by showing how the states of all three modes in a DHD crystal can be transformed through modulation of the trapping potential while the ions separate to allow for GGS in a three ion crystal without prior or posterior operations.  We find that realistically feasible potential modulation is sufficient to end near the ground states of all three modes and that the protocol is robust against realistic levels of imperfection of the modulation. Concluding remarks in Sec.~\ref{sec:Conclusion} summarize the topics introduced in this paper, compare this work to other proposals, and discuss opportunities for future work.
\section{Theoretical Formalism} \label{sec:Formalism}
Before getting into the specifics of any particular ion chain or transport protocol, we will review a theoretical formalism which describes the quantum dynamics of systems undergoing time evolution through quadratic Hamiltonians.  The formalism itself is general, however we will only apply it to Gaussian states. More details can be found in \cite{Simon1994,Weedbrook_2012,Hackl_2018,brask2022gaussian}.
\subsection{Evolution under quadratic Hamiltonians}
We denote the generalized time-independent momentum and position operators of $N$ particles in one dimension as
\begin{align}
    \xi &= (p_1, \cdots, p_N, x_1, \cdots, x_N)^T. 
\end{align}
In our context these are the momentum and position operators for normal modes of coupled ion motion of $N$ ions along the axis of a linear crystal. Components $\xi_a$ and $\xi_b$ satisfy the following commutation relations (note that from here on $\hbar$ has been set to $1$):
\begin{align}
    \comm{\xi_a}{\xi_b} &= i C_{ab},\\
    \boldsymbol{C} &= \begin{pmatrix}
        \boldsymbol{0} & -\mathbb{I} \\
        \mathbb{I} & \boldsymbol{0}
    \end{pmatrix},
\end{align}
with $\mathbb{I}$ the $N$-dimensional identity matrix.\\
\\
We further assume that the system evolves under a time dependent purely quadratic Hamiltonian that can be written as
\begin{align}
    H(t) &= \frac12 \xi^T \boldsymbol{h}(t) \xi, \label{Eq:generalHam}
\end{align}
where $\boldsymbol{h}(t)$ is a Hermitian matrix with dimension $2N\times2N$. Generalization to include sub-quadratic terms is straightforward and not important for this work. We can now write down how $\xi$ evolves in the Heisenberg picture.  Going forward, $\Tilde{\xi} = (\Tilde{p}_1,\cdots,\Tilde{p}_N, \Tilde{x}_1, \cdots, \Tilde{x}_N)^T$, will represent time-dependent Heisenberg picture operators.  Formally, we can write the time evolution as the time-ordered exponential
\begin{align}
    U(t) &= \mathcal{T}\exp(-i \intl_0^t H(t') dt')\\
    \dot{\tilde{\xi}}_a(t) &= i U\dagg \comm{H(t)}{\xi_a} U.
\end{align}
Hence,
\begin{align}
    \dot{\tilde{\xi}}(t) &= \boldsymbol{C}\cdot \boldsymbol{h}(t) \tilde{\xi}(t),
    \label{Eq:XiEqMotion}
\end{align}
which is equivalent to the classical equations of motion for the momentum and position coordinates, $\Xi = (P_1, \cdots, X_1, \cdots)$.  This equation can be rewritten by defining the $2N\times2N$ transfer matrix $\boldsymbol{M}(t)$,
\begin{align}
    \tilde{\xi}(t) &= \boldsymbol{M}(t) \tilde{\xi}_0,
    \label{eq:MDefinition}
\end{align}
which takes $\tilde{\xi}(t=0) = \tilde{\xi}_0$ to some later time $t$. Inserting into (\ref{Eq:XiEqMotion}) yields
\begin{align}
    \dot{\boldsymbol{M}}(t) &= \boldsymbol{C}\cdot 
    \boldsymbol{h}(t)\cdot \boldsymbol{M}(t). \label{eq:MSolution}
\end{align}
With a solution to (\ref{eq:MSolution}) in hand, the dynamics of $\tilde{\xi}(t)$ are fully determined.  From those dynamics, the first and second moments of $\tilde{\xi}$ are also determined.  When restricted to Gaussian states, the covariance matrix of $\tilde{\xi}$, $\boldsymbol{V}$, evolves as \cite{Simon1994,Weedbrook_2012,Hackl_2018}
\begin{align} 
    \boldsymbol{V}(t) &= \boldsymbol{M}(t)\cdot\boldsymbol{V}_0 \cdot \boldsymbol{M}^T(t).
     \label{eq:covarEvolve}
\end{align}
This can be understood by realizing that the Wigner function of a Gaussian state evolving under a quadratic potential is constant along classical trajectories \cite{Schleich}:
\begin{align}
    W(\Xi, t) = W_0(\Xi_0(\Xi, t), t = 0).
\end{align}
Hence, the quadratic form appearing in the initial Gaussian Wigner function evolves as
\begin{align}
\Xi_0^T \boldsymbol{V}_0^{-1} \Xi_0 &\to \Xi^T [\boldsymbol{M}^{-1}(t)]^T \cdot\boldsymbol{V}_0^{-1} \cdot \boldsymbol{M}^{-1}(t)\Xi\, ,
\end{align}
which then implies (\ref{eq:covarEvolve}).  As such, the state is fully specified for the particular case of Gaussian states \cite{Schrade_1995,Schleich}.  This will be important later on when we evolve the ground state of a linear ion crystal.  Another important factor to consider is that $\boldsymbol{M}\cdot \boldsymbol{C} \cdot\boldsymbol{M}^T = \boldsymbol{C}$, which implies $\boldsymbol{M}$ is symplectic, and parameterizes $Sp(2N,\mathbb{R})$, the group under which dynamical time evolution occurs \cite{SudarshanSimon,Simon1994,SIMON1987223,Arvind_1995,Glauber,Hackl_2018}.

The matrix $\boldsymbol{M}(t)$ can be Bloch-Messiah decomposed into a combination of multi-mode interferometers $\boldsymbol{B}(\boldsymbol{\theta}_k)$ and single-mode squeezers $\boldsymbol{S}(r_k,\phi_k)$ as \cite{BLOCH196295,Braunstein2005,Simon1994,Weedbrook_2012}
\begin{align} \label{Eq:EulerDecomp}
    \boldsymbol{M}(t) &= \boldsymbol{B}(\boldsymbol{\theta}_2)\cdot \left[\bigoplus_{k=1}^N \boldsymbol{S}(r_k,\phi_k) \right] \cdot \boldsymbol{B}(\boldsymbol{\theta}_1),
\end{align}
where $\boldsymbol{\theta}_k$, $\phi_k$ and $r_k$ will in general be time dependent, and $\boldsymbol{\theta}_k$ are Hermitian matrices composed of rotation angles and mixing angles that determine couplings between different modes. The interferometers and squeezers have the functional form
\begin{align}
    \boldsymbol{B}(\boldsymbol{\theta}) &= e^{i \boldsymbol{a}\dagg\cdot \boldsymbol{\theta} \cdot \boldsymbol{a}}\\
    \boldsymbol{S}(r_k, \phi_k) &= \exp[\frac{r_k}{2}\left((a_k^\dagger)^2 e^{i\phi_k} - a_k^2 e^{-i\phi_k}\right)],
\end{align}
where $\boldsymbol{a} = (a_1,\dots,a_N)$ and the ladder operators of mode $k$ are denoted as $a_k$ and $a_k^\dag$. For two modes, a two-port interferometer is equivalent to the application of a beam-splitter and a rotation \cite{Reck1994,Clements:16} 
\begin{align} \label{eq:twoModeInter}
    \boldsymbol{B}(\boldsymbol{\theta}) &= \boldsymbol{B}_{BS}(\theta_{BS}, \phi_{BS}) \cdot \bigg[\boldsymbol{R}_1(\theta_{1}) \oplus \boldsymbol{R}_2(\theta_{2}) \bigg]\\
    &\equiv \boldsymbol{B}_{BS}(\theta_{BS}, \phi_{BS}) \cdot \boldsymbol{R}_{12}(\theta_{1}, \theta_2)
\end{align}
with
\begin{align}
    &\boldsymbol{B}_{BS}(\theta_{BS}, \phi_{BS}) = \exp[\theta_{BS} \left(a_a\dagg a_b e^{i\phi_{BS}} - a_a a_b\dagg e^{-i\phi_{BS}} \right)],
\end{align}
and
\begin{align}\label{Eq:Rotation}
    \boldsymbol{R}_k(\theta_k) &= \exp[-i\theta_k a_k\dagg a_k]\\
    \boldsymbol{R}_{lm}(\theta_l, \theta_m) &= \boldsymbol{R}_l(\theta_{l}) \oplus \boldsymbol{R}_m(\theta_{m})
\end{align}
A detailed explanation on how this decomposition in (\ref{Eq:EulerDecomp}) is performed can be found in \cite{cariolaro2016}.

Crucially, this tells us that the dynamics of any time-dependent quadratic Hamiltonian can be decomposed as a combination of single-mode squeezers and beam-splitting operations. This enables efficient computation of suitable further squeezing and beam-splitting operations, either before or after $H(t)$ acts, that compensate for the effects of $\boldsymbol{M}(t)$ on the initial ground states and will result in final ground states, generally in a basis that can be different from the initial phase space basis. Formally, if $\boldsymbol{Q}$ is the orthogonal matrix that transforms the initial operators $\tilde{\xi}_0$ that diagonalize $H(0)$ into a set of operators $\tilde{\xi}_t$ that diagonalize $H(t)$, then for GGS
\begin{align} 
    \boldsymbol{T} \cdot \boldsymbol{M}(t) = \boldsymbol{M}(t) \cdot \boldsymbol{T}' =\boldsymbol{Q},
    \label{Eq:CompensationMatrices}
\end{align}
where $\boldsymbol{T}$ ($\boldsymbol{T}'$) are suitable compensation operations composed of single mode squeezers, multi-port interferometers and rotations applied after (before) $H(t)$ has acted. Special cases of this general principle  for a single normal mode of ion motion, and for two decoupled normal modes were explored in \cite{Sutherland:2021ifb}. We will discuss the more general separation of a DHD crystal in more detail in section \ref{sec:TransProt}. Alternatively, a specific $\boldsymbol{h}(t)$ can be identified such that no initial or final operation is necessary for GGS,
\begin{align} 
    \boldsymbol{M}(t) = \boldsymbol{Q},
    \label{Eq:GtoG}
\end{align}
which removes the necessity to compensate for the effects of $H(t)$ and potentially reduces the total duration of all operations necessary for GGS. This approach to separation of a DHD crystal is discussed in section \ref{sec:fastThreeIon}.\\  
\\
In both cases, since $\boldsymbol{M}(t)$ represents a unitary time evolution of the system, the equations of motion are invariant under time reversal. As long as this is the case, the operators $\boldsymbol{M}(t), \boldsymbol{T}$ and $\boldsymbol{T}'$ can be inverted to run the process in reverse. This implies that a good solution for ion separation also yields a good solution for the time reversed process, ion recombination.
\subsection{Occupation Numbers from the Covariance Matrix}
Given a quadratic Hamiltonian, we would like to track various mode occupation numbers over time.  For example, if the curvature of a potential well is changing in time, we may be interested in tracking the occupation numbers of the normal modes of ions confined in the well.  In the previous section, we explained how the covariance matrix of a Gaussian state evolves in time under a quadratic Hamiltonian. The occupation number for a mode $k$ with frequency $\omega_k$ that remains uncoupled from all other modes at all times has expectation value
\begin{align}
    \expval{n_k} &= \frac{\expval{H_k(t)}}{\omega_k} - \frac12,
\end{align}
where $H_k(t)$ is the decoupled subsystem's Hamiltonian. The Hamiltonian's expectation value can be computed from the isolated block in the covariance matrix $\boldsymbol{V}$ of the whole system corresponding to mode $k$ as
\begin{align}
    \expval{H_k(t)} &= \frac12 V_{11}(t) + \frac12 \omega_k^2 V_{22}(t),
\end{align}
where the matrix subscripts indicate the row and column of the covariance matrix element. In general, when modes are coupled they do not have a well defined occupation number.  However, if a mode is decoupled at any time (for example at late times in a separation protocol, when the ions are far apart from each other), we can extend the definition at that time to all times to define a quantity of interest.  A natural choice of modes are the final modes of a separated crystal. After separating an ion crystal into individual potential wells, the final modes will be completely decoupled as the Coulomb repulsion between ions will be negligible, and each ion can be considered individually.  Throughout this paper, we will make the choice to compute occupation number with respect to the final modes of a DHD crystal.  For example, if two modes, $j$ and $k$, are coupled initially, but decoupled at late times with potential strengths $\omega_{j,k}$, we can define the following quantities derived from their covariance matrix $\boldsymbol{V}^{jk}$ at all times,
\begin{align}
    \expval{n_j} &= \frac{\frac12 V^{jk}_{11}(t) + \frac12 \omega_j^2 V^{jk}_{33}(t)}{\omega_j} - \frac12 \\
    \expval{n_k} &= \frac{\frac12 V^{jk}_{22}(t) + \frac12 \omega_k^2 V^{jk}_{44}(t)}{\omega_k} - \frac12.
\end{align}
This can be generalized to $M$ coupled modes that are uncoupled at some time, but the dimension of the block of correlations in the covariance matrix of the full system that needs to be considered increases to $2M \times 2M$. 
\section{Two-Species Three-Ion  (DHD) Crystal} \label{sec:TransProt}
We will now show how GGS can be performed in a linear DHD crystal, which we treat in one spatial dimension $x$ along the crystal axis.  The specific ion species under consideration are unimportant for this work, however for specificity when performing numerical calculations we choose the mass $m_D$ of D equal to that of \ion{Be}{9} and $m_H$ of H equal to that of \ion{Mg}{25}. 

In the separation considered here, we assume that a linear DHD crystal is aligned along the weakest axis of the trapping potential and its axial normal modes are cooled to near their ground states, with the expectation values of the axial ion positions equal to the classical equilibrium positions of the ions denoted by $c_{D1}$, $c_{D2}$, and $c_H$, which will become time-dependent during the separation protocol. The H species ion is held in between the two D species ions. The applied harmonic confining potential at the ion positions is defined in terms of a local spring constant, $k_H(t)$ at the H ion and $k_D(t)$ at the D ions. When this potential is removed, the two D ions are pushed apart by the three ions' mutual Coulomb repulsion, while the H ion ideally remains stationary at the origin at all times. After some amount of time, the D ions have reached a sufficient distance such that approximately harmonic potentials local to each ion can be turned back on with negligible interference between the three wells, thus trapping each ion individually.  When neglecting the residual Coulomb repulsion at the final ion distances, the separate potential well minimum locations $w_{D1}$ and $w_{D2}$ coincide with the classical positions $c_{D1}$ and $c_{D2}$ of the D ions, symmetrically displaced from the origin where the H ion remained during the separation.\\
\\
This can be described by the classical Hamiltonian 
\begin{align}
    \mathcal{H}(t) &= \frac{p_{D1}^2 + p_{D2}^2}{2m_D} + \frac{p_H^2}{2m_H} + \frac12 k_D(t) \left[ (c_{D1} - w_{D1})^2 + (c_{D2} - w_{D2})^2\right] + \frac12 k_H(t) c_H^2  \nonumber \\
    &+  \frac{k_e}{c_H - c_{D1}} +  \frac{k_e}{c_{D2} - c_{D1}} +  \frac{k_e}{c_{D2} - c_H}.
    \label{Eq:FullHam}
 \end{align}
where $p_{D1}$, $p_{D2}$, and $p_{H}$ are the momenta of the three ions. We note that the $c$ and $w$ values are time dependent although we have not written this explicitly in ~\ref{Eq:FullHam}. The DHD crystal is arranged such that $c_{D1}<c_H=0<c_{D2}$ so all denominators in the Coulomb interaction terms are positive throughout the separation. The constant $k_e = q^2/(4 \pi \epsilon_0)$ with $q$ the elementary charge and $\epsilon_0$ the vacuum permittivity scales the Coulomb interaction.  We make the approximation that the classical position of the H ion remains fixed, $c_H(t) = 0$, and do not consider perturbations of its position at this point.  This approximation also implies $c_{D2}(t) = -c_{D1}(t) \equiv c(t)$ and $w_2(t) =-w_1(t)=w(t)$, with $w_H(t)=0$.\\
\\
To model the small oscillations of the ions around their classical motion quantum mechanically, we can transform into a frame of reference moving along the classical trajectory $c_{D1}(t)$, $c_{D2}(t)$, and $c_{H}(t)$ for each ion (determined by solving~\ref{Eq:FullHam}) by applying appropriate displacement operators \cite{Sutherland:2021ifb}. We can then introduce mass weighted quantum mechanical operators 
\begin{align}
    p_j &\to \sqrt{m_j} p_j, \hspace{.1\linewidth}
    x_j \to \frac{x_j}{\sqrt{m_j}}.
\end{align}
to describe small displacements relative to the classical frame of reference and accommodate ions of different mass. Under the assumption that the real space (non mass weighted) displacements are much smaller than the relative distances between ions, we can also expand the Coulomb term to quadratic order. Reinterpreting the classical position and momentum variables as quantum mechanical operators, we arrive at a quantum mechanical Hamiltonian corresponding to three coupled harmonic oscillators:
\begin{align}
    H(t) &\approx \frac{1}{2}\left[ p_{D1}^2 + p_{D2}^2+p_H^2\right] + \frac12 \frac{k_D(t)}{m_D} \left[ x_{D1}^2 + x_{D2}^2 \right] + \frac{1}{2} \frac{k_H(t)}{m_H}x_H^2\nonumber\\
    &+ \frac{k_e}{c^3(t)}\left[\frac{x_{D1}^2+x_{D2}^2}{m_D} + \frac{2 x_{H}^2}{m_H} + \frac{(x_{D2} - x_{D1})^2}{8 m_D}  - \frac{2 x_H(x_{D1}+x_{D2})}{\sqrt{m_D m_H}}\right].
   \label{Eq:QuadHam}
\end{align}
As long as the separating ions are sufficiently closely approximated by (\ref{Eq:QuadHam}), the general formalism described in section \ref{sec:Formalism} can be applied to describe the ensuing dynamics.\\
\\
We can partially decouple the oscillators by introducing in-phase and out-of-phase coordinates for the D ions,
\begin{align}
    x_{op} &= \frac{x_{D2} - x_{D1}}{\sqrt{2}}, \hspace{.05\linewidth}
    &&x_{ip} = \frac{x_{D2} + x_{D1}}{\sqrt{2}}\\
    p_{op} &= \frac{p_{D2} - p_{D1}}{\sqrt{2}}, \hspace{.05\linewidth}
    &&p_{ip} = \frac{p_{D2} + p_{D1}}{\sqrt{2}},
\end{align}
along with corresponding time-dependent oscillator frequencies
\begin{align}
      \omega_{op}^2(t) = \frac{k_D(t)}{m_D} + \frac{5 k_e}{2m_D c^3(t)},\nonumber\\
      \omega_{ip}^2(t) = \frac{k_D(t)}{m_D} + \frac{2k_e}{m_D c^3(t)},\nonumber\\ 
      \omega_H^2(t) = \frac{k_H(t)}{m_H} + \frac{4k_e}{m_H c^3(t)},
       \label{Eq:OscFreq}
    \end{align}
and mode coupling strength
\begin{align}
       \Omega_{Hip}^2(t) = \frac{4\sqrt{2}~k_e}{\sqrt{m_Dm_H} ~c^3(t)}.
\end{align}
The Hamiltonian can then be written as
    \begin{align}
        H(t) &= \frac12 \sum_{k=ip,op,H} \left[p_k^2 + \omega_k^2(t) x^2_k\right] - \Omega_{Hip}^2(t) x_Hx_{ip}, \label{Eq:secondTrans}
    \end{align}
which has the form of the Hamiltonian in (\ref{Eq:generalHam}) with an out-of-phase mode that is decoupled from the other two modes and has no participation of the H ion. The other two modes are coupled initially with a term that is linear in their position operators and therefore acts in analogy to a beam-splitter. Inspection of (\ref{Eq:OscFreq}) reveals that the terms proportional to the Coulomb interaction fall off as $1/c^3(t)$ and rapidly become negligible compared to the confinement $k_D(t)$ from the external potential with increasing $c(t)$. This justifies the approximation of three decoupled oscillators at the end of separation with motional frequencies determined entirely by the local curvature of the external potential and the mass of the ions.\\
\\
Applying a suitable time-dependent basis rotation between the two coupled modes will allow us to write the Hamiltonian in a way that is fully decoupled whenever the oscillator frequencies are not changing in time (in particular, before and after separation).  However, the modes can still be coupled at intermediate times. To this end, we can define two new normal modes, $a$ and $b$, as well as their operators, $x_a$ and $x_b$ that are connected to the H and $ip$ modes by the unitary transformation 
\begin{align}
U_\theta = \exp\left[-i\theta \left(p_H x_{ip} - p_{ip} x_H\right)\right]
  \end{align}
where $\theta$ is implicitly defined as 
\begin{align}
    \tan 2 \theta &= \frac{2 \Omega_{Hip}^2}{\omega_H^2 - \omega_{ip}^2} \label{eq:thetaDef}.
\end{align}
The position operators in the rotated basis are,
\begin{align}
		x_a = U_\theta x_{ip} U_\theta\dagg &= x_{ip} \cos\theta + x_H\sin\theta, \nonumber\\
		x_b = U_\theta x_H U_\theta\dagg &= x_H \cos\theta - x_{ip}\sin\theta.
\end{align}
Since $\theta$ changes in time, this basis is constantly adjusting as the separation proceeds. In the adjusted basis, the Hamiltonian transforms to
\begin{eqnarray}
    H_\theta &=& U_\theta H(t) U_\theta\dagg + i \dot{U}_\theta U_\theta\dagg = \frac{p_{ip}^2 + p_a^2 + p_b^2}{2} + \frac12 \omega_{ip}^2(t) x_{ip}^2 + \nonumber\\
    &&\frac12 \omega_a^2(t) x_a^2 + \frac12 \omega_b^2(t) x_b^2 + \dot{\theta} \left(p_bx_a - p_ax_b\right), \label{Eq:decoupHam}
\end{eqnarray}
where we have defined the following quantities,
\begingroup
\allowdisplaybreaks
\begin{align}
    \omega_a^2 &= \frac12 (\omega_{ip}^2 + \omega_H^2 + \Gamma)\\
    \omega_b^2 &= \frac12 (\omega_{ip}^2 + \omega_H^2 - \Gamma)\\
    \Gamma &= \sqrt{4\Omega_{Hip}^4 + (\omega_{ip}^2 - \omega_H^2)^2}.
\end{align}
\endgroup
A full exploration of this type of transformation in two dimensions can be found in \cite{Lizuain_2017}. In the chosen frame of reference, all modes are decoupled whenever $\dot{\theta} = 0$. Additionally, the position of one oscillator is coupled to the momentum of the other in this frame, which can be seen from the last term in (\ref{Eq:decoupHam}). 
%
\subsection{Classical Dynamics}
For separation of the DHD crystal in a physical system, the applied confining potential cannot be turned on and off instantaneously due to experimental constraints. For concreteness and ease of calculation in the following examples, we  use sinusoidal ramping of the potential in time, but other ramp shapes could also be used. We ramp the external axial potential to zero starting at $t=0$ over a duration $\tau$, followed by expansion of the ion crystal for $\tau_0$ and another sinusoidal ramp to re-trap the ions in separate wells with duration $\tau$, in analogy to the separation considered in \cite{Sutherland:2021ifb}. Writing out the time dependence explicitly and denoting $\omega_0 =\sqrt{k_D(0)/m_D}$, 
\begin{align}
    \omega(t) =
    \begin{cases}
        \omega_0 & t \le 0\\
        \frac{\omega_0}{2} \left[ 1+\cos\left(\frac{\pi}{\tau}t\right)\right] & 0 < t \le \tau \\
        0 & \tau < t \le \tau + \tau_0 \\
        \frac{\omega_0}{2} \left[ 1-\cos(\frac{\pi (t-\tau-\tau_0)}{\tau})\right] & \tau + \tau_0 < t \le 2\tau +\tau_0 \\
        \omega_0 & t > 2\tau +\tau_0. \label{eq:potentialShape}
    \end{cases}
\end{align}
It is necessary to ensure that the classical motion of all three ions leaves them at rest at the end of the separation. While the initial potential confines the ions, $t\in (-\infty, \tau]$, the potential minimum is located at the origin, $w_{i}(t < \tau) = 0$.  However, during the re-trapping period, we will allow the potential to move and apply forces to the D ions to slow them down.  To this end, the potential minima near the D ions will follow
\begin{align}\label{Eq:friction_term}
    w_{1}(t) = x_{D1}(t) - \eta \dot{x}_{D1}(t)
\end{align}
and analogously for $w_{2}$ with $x_{D2}$. This introduces decelerating forces proportional to the ion velocities in the classical dynamics which slow the ions as they are recaptured and cease once the ions are at rest.
\subsection{State Preparation} \label{sec:StatePrep}
To achieve GGS, we can cool the motional modes close to their ground states and pre-compensate for the quantum mechanical effects of separation during $t<0$, then separate the crystal starting at $t=0$ to end in the ground states of all three final modes at $t_f$. We can separate (\ref{Eq:decoupHam}) into the time evolution of two states. The $ip$ mode is completely decoupled from the rest of the Hamiltonian and can be evolved independently. The $a$ and $b$ modes are decoupled at $t=0$, couple during separation (implying that their motion is entangled) and decouple at $t=t_f=2 \tau+\tau_0$. They must be evolved together. The time evolution during separation can be further decomposed into suitable squeezing operations $\boldsymbol{S}(r_k,\phi_k)$ on the $k$-th initial normal mode with ($k=\{op,a,b\}$), rotation operations $\boldsymbol{R}(\theta_k)$ on the $k$-th initial normal mode and a beam-splitting operation $\boldsymbol{B}_{BS}(\theta_{BS}, \phi_{BS})$ between modes $a$ and $b$ as defined in (\ref{Eq:EulerDecomp})-(\ref{Eq:Rotation}). In the phase space coordinates introduced in \ref{sec:Formalism}, these operations take the form 
{\small
\begin{align}
    \begin{pmatrix}
        p_k \\
        x_k
    \end{pmatrix}_S
    &= \boldsymbol{S}(r_k, \phi_k) \begin{pmatrix}
        p_k\\
        x_k
    \end{pmatrix}_0
    = \begin{pmatrix}
        \cosh(r_k) - \sinh(r_k)\cos(\phi_k) & \omega_k \sinh(r_k)\sin (\phi_k) \\
        \frac{1}{\omega_k}\sinh(r_k)\sin(\phi_k) & \cosh(r_k) + \sinh(r_k)\cos(\phi_k)
    \end{pmatrix}
    \begin{pmatrix}
        p_k\\
        x_k
    \end{pmatrix}_0 \label{Eq:singleSqueeze} \\ 
    \begin{pmatrix}
        p_k \\
        x_k
    \end{pmatrix}_R
    &= \boldsymbol{R}(\theta_k) \begin{pmatrix}
        p_k\\
        x_k
    \end{pmatrix}_0
    = \begin{pmatrix}
        \cos\theta_k & -\omega_k\sin\theta_k \\
        \frac{1}{\omega_k}\sin\theta_k & \cos\theta_k
    \end{pmatrix}
    \begin{pmatrix}
        p_k\\
        x_k
    \end{pmatrix}_0 \label{Eq:singleRotation}
\end{align}
}
{\scriptsize
\begin{align}
    \begin{pmatrix}
        p_a\\
        p_b\\
        x_a\\
        x_b
    \end{pmatrix}_{BS}
    = \begin{pmatrix}
        \cos\theta_{BS} & \sqrt{\frac{\omega_a}{\omega_b}} \cos(\phi_{BS})\sin(\theta_{BS}) & 0 & \sqrt{\omega_a \omega_b} \sin(\phi_{BS})\sin(\theta_{BS}) \\
        - \sqrt{\frac{\omega_b}{\omega_a}} \cos(\phi_{BS})\sin(\theta_{BS}) & \cos\theta_{BS} & \sqrt{\omega_a \omega_b} \sin(\phi_{BS})\sin(\theta_{BS}) & 0 \\
        0 & \frac{-1}{\sqrt{\omega_a\omega_b}} \sin(\phi_{BS})\sin(\theta_{BS}) & \cos\theta_{BS} & \sqrt{\frac{\omega_b}{\omega_a}} \cos(\phi_{BS})\sin(\theta_{BS}) \\
        \frac{-1}{\sqrt{\omega_a\omega_b}} \sin(\phi_{BS})\sin(\theta_{BS}) & 0 & - \sqrt{\frac{\omega_a}{\omega_b}} \cos(\phi_{BS})\sin(\theta_{BS}) & \cos\theta_{BS}
    \end{pmatrix}
    \begin{pmatrix}
        p_a\\
        p_b\\
        x_a\\
        x_b
    \end{pmatrix}_0
\end{align}
}
where $\omega_k =\omega_k(0)$, and $\omega_{a,b} =\omega_{a,b}(0)$. Writing the operators in this form requires application of the Baker-Campbell-Hausdorff theorem, and is covered in detail in \cite{brask2022gaussian}. The arguments for these operations are determined by the quantum dynamics of the system during separation. In the frame moving with the classical ion positions, these are governed by (\ref{Eq:decoupHam}), which we will consider in detail in the next section. To reasonably compensate for the beam-splitting dynamics and squeezing during separation, all three ions need to be in close proximity with substantial Coulomb coupling and resolvable differences in the mode frequencies. For this reason, we will only consider compensation before or during separation here.  
\subsection{Time Evolution and Occupation Numbers}
\subsubsection{The out-of-phase mode}
To map an arbitrary state in the initial well at frequency $\omega_{op}(0)=\sqrt{3} \omega_0$ to its equivalent state in the final well at frequency $\omega_{op}(t_f)=\omega_0$ the net effect of all operations must be equal to the scaling operator 
\begin{align}
\boldsymbol{Q}_{op} = 
   \begin{pmatrix} 
        \sqrt{\frac{\omega_{op}(t_f)}{\omega_{op}(0)}} & 0 \\
       0 & \sqrt{\frac{\omega_{op}(0)}{\omega_{op}(t_f)}}
    \end{pmatrix},
\end{align}
which can also be thought of as undoing the effect of the squeezing incurred by the out-of-phase mode due to an instantaneous change in its frequency from $\omega_{op}(0)$ to $\omega_{op}(t_f)$.\\
\\
In terms of the covariance, the out-of-phase mode's initial ground state at the initial oscillator frequency, $\sqrt{3} \omega_0$, is described by
\begin{align}
    \boldsymbol{V}_{op}(0) &\equiv \boldsymbol{V}^{(i)}_{op} =
    \begin{pmatrix}
        \frac{\sqrt{3} \omega_0}{2} & 0 \\
        0 & \frac{1}{2 \sqrt{3} \omega_0}
    \end{pmatrix},
\end{align}
while the covariance matrix of the out-of-phase modes's final state after successful GGS is given by
\begin{align}
    \boldsymbol{V}_{op}(t_f) &\equiv \boldsymbol{V}^{(f)}_{op} =
    \begin{pmatrix}
        \frac{\omega_0}{2} & 0 \\
        0 & \frac{1}{2\omega_0}
    \end{pmatrix},
\end{align}
and $\boldsymbol{Q}_{op}\cdot \boldsymbol{V}^{(i)}_{op}\cdot \boldsymbol{Q}^T_{op}=\boldsymbol{V}^{(f)}_{op}$ as expected.\\
\\
The effect of separation on the out-of-phase mode can be calculated from $\boldsymbol{h}(t)$ in (\ref{Eq:generalHam})
\begin{align}
    \boldsymbol{h}_{op}(t) &= 
    \begin{pmatrix}
        1 & 0 \\
        0 & \omega_{op}^2(t)
    \end{pmatrix},
\end{align}
which can be inserted into  (\ref{eq:MSolution}) to solve for $\boldsymbol{M}_{op}(t)$ numerically
\begin{align}
    \dot{\boldsymbol{M}}_{op}(t) &= - \boldsymbol{C}_{op} \cdot \boldsymbol{h}_{op}(t) \cdot \boldsymbol{M}_{op}(t) \nonumber\\
    \boldsymbol{M}_{op}(0) &= \boldsymbol{I},
\end{align}
to yield $\boldsymbol{M}_{op}(t_f)\equiv \boldsymbol{M}^{(f)}_{op}$. The Bloch-Messiah theorem states that $\boldsymbol{M}^{(f)}_{op}$ can be decomposed into squeezing operations and rotations of the form given in (\ref{Eq:singleSqueeze}) and (\ref{Eq:singleRotation}), each with unit determinant. Therefore, the inverse is
\begin{align}
  (\boldsymbol{M}^{(f)}_{op})^{-1} &= 
  \begin{pmatrix}
        m_{22} & -m_{12} \\
        -m_{21} & m_{11}
    \end{pmatrix},
\end{align}
where $m_{jk}$ are the matrix elements of $ \boldsymbol{M}^{(f)}_{op}$. The covariance $\boldsymbol{V}^{(p)}_{op}$ at $t=0$, after pre-compensation but before separation is
\begin{align} \label{Eq:precovariance}
    \boldsymbol{V}^{(p)}_{op}\equiv(\boldsymbol{M}^{(f)}_{op})^{-1} \cdot \boldsymbol{V}_f \cdot \left[(\boldsymbol{M}^{(f)}_{op})^{-1}\right]^T.
\end{align}
The general form of a pre-compensation operation is
\begin{align}\label{Eq:Mpdecomp}
    \boldsymbol{M}^{(p)}_{op} &=\boldsymbol{R}_p(\theta_{2})\cdot \boldsymbol{S}_p(r,\phi)\cdot \boldsymbol{R}_p(\theta_{1})\nonumber\\
    &=\boldsymbol{R}_p(\theta_{2})\cdot \boldsymbol{S}_p(r,\phi)\cdot \boldsymbol{R}_p(-\theta_{2})\cdot \boldsymbol{R}_p(\theta_{2})\cdot \boldsymbol{R}_p(\theta_{1})\nonumber\\
    &= \boldsymbol{S}_p(r_p,\phi_p)\cdot\boldsymbol{R}_p(\theta),
\end{align}
where $\theta_p = \theta_{1}+\theta_{2}$, and $r_p$ and $\phi_p = \phi + 2 \theta_{2}$ are parameters that are not yet determined. Since $\boldsymbol{V}_0$ is the covariance of a Gaussian state, $\boldsymbol{R}_{p} (\theta) \cdot \boldsymbol{V}_p(0) \cdot \boldsymbol{R}^T_{p} (\theta)=\boldsymbol{V}_{op}(0)$ for any $\theta$. After pre-compensation, the covariance matrix of the initial state takes the form
\begin{align} \label{Eq:symmetricSolve}
    \boldsymbol{V}^{(p)}_{op}&= \boldsymbol{S}_p(r_p, \phi_{p}) \cdot \boldsymbol{V}^{(i)}_{op} \cdot \boldsymbol{S}^T_p(r_p, \phi_{p}).
\end{align}
By inserting (\ref{Eq:singleSqueeze}) into (\ref{Eq:symmetricSolve}) and comparing to (\ref{Eq:precovariance}) element by element we find
\begin{align}
    \cosh({2r_p}) &= \frac{\left(\frac{m_{12}}{\omega_0}\right)^2 + 3 m_{11}^2 + \left(m_{22}^2 + 3 m_{21}^2 \omega_0^2\right)}{2\sqrt{3}} \\
    \sin(\phi_p) &= - \frac{m_{11}m_{12} + m_{22}m_{21} \omega_0^2}{\omega_0 \sinh{2r_p}}.
\end{align}
For a concrete example, we assume a BMB crystal with initial axial in-phase frequency $\omega_{ip}/2\pi = 1$~MHz, where $\omega^2(t)$ is ramped down over $\tau = 0.365~\mu$s.  The ions are allowed to fly apart unimpeded for $\tau_0 = 1.1 ~\mu$s, and the potentials are ramped back up over $2 \tau = 0.73 ~\mu$s. The required pre-squeezing parameters are $r_p \approx 1.597$, and $\phi_{p} \approx -0.671$. 
\begin{figure}[h!]
\includegraphics[width=3.5in, height=3.5in]{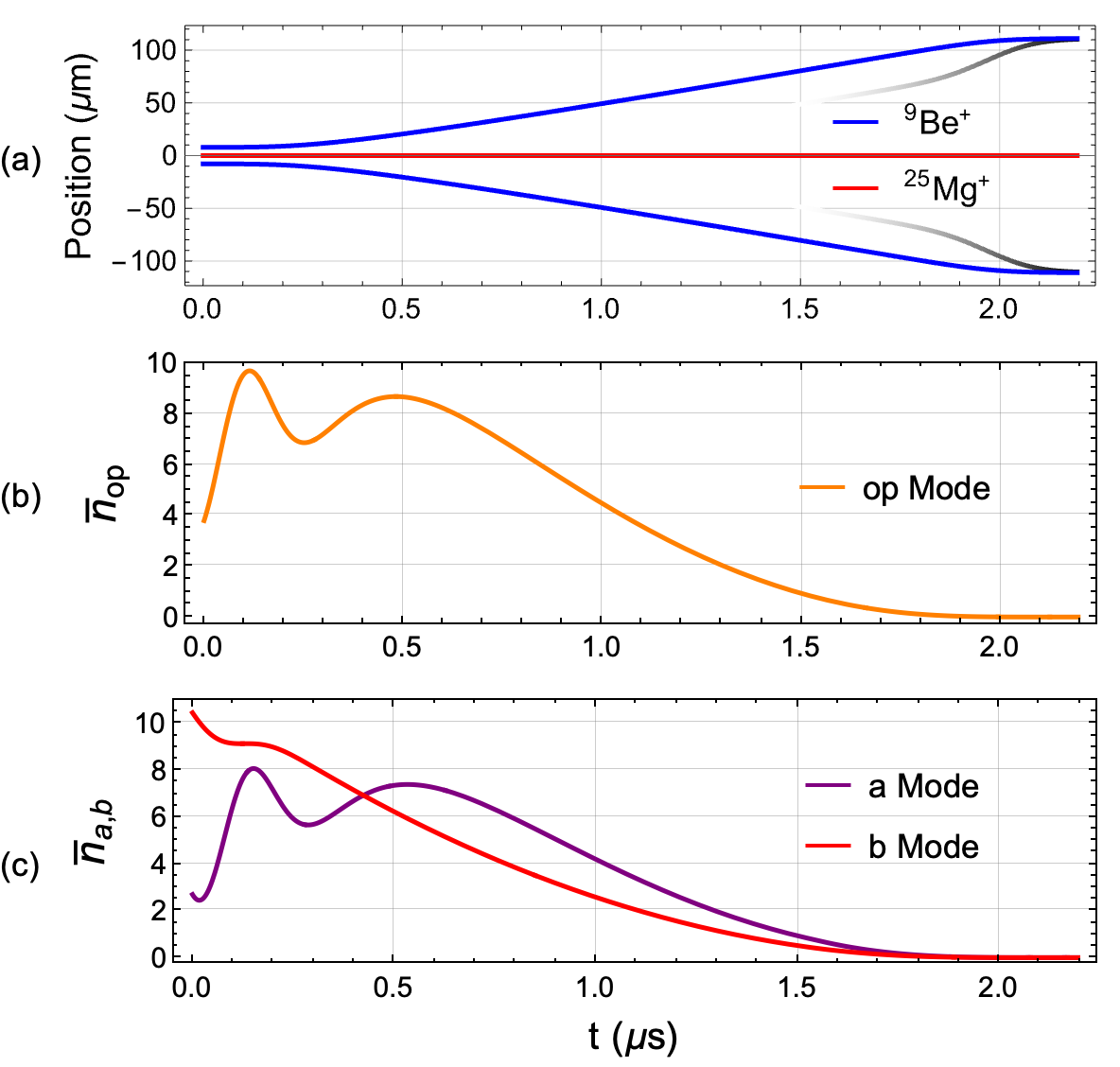}
\caption{Two numerical solutions. (a) Classical time evolution of the three ion BMB crystal.  At $t=0$ the trapping potential is ramped down from the equilibrium strength over $0.365 ~\mu $s.  At $t=1.465 ~\mu $s the potential begins ramping up over $0.73 ~\mu$s.  This amounts to a total separation time of $\sim2.2 ~\mu$s without accounting for state preparation. The grayscale line indicates individual trapping potential locations for the B ions as well as strength, with higher opacity indicating stronger confinement.  The M ion ideally remains in place with its confining potential strength ramped simultaneously with those for the B but not shown in the figure. (b) Mean occupation number of the $op$ mode during time evolution. (c) Mean occupation numbers of the $a$ and $b$ modes during time evolution.} \label{fig:timeEvolvePlots}
\end{figure}
The total duration of separation is similar to the two-ion separation discussed in \cite{Sutherland:2021ifb}, which is reasonable given that similar Coulomb repulsion forces, ion masses, and final relative distances are used.\\
\\
The correct parameters for pre-compensation and the numerical solution for $\boldsymbol{M}_{op}(t)$ allows us to compute the occupation number of the out-of-phase mode as a function of time with respect to the final well frequency $\omega_0$ from $t=0$ to $t_f$,
\begin{align}
    \expval{n_{op}}_t &= \frac{\frac12 V^{op}_{11}(t) + \frac12 \omega_0^2 V^{op}_{22}(t)}{\omega_0} - \frac12.
\end{align}

Fig \ref{fig:timeEvolvePlots} shows (a) the position and strength of the potential wells and the classical time evolution of the position of the three ions and (b) the out-of-phase mode's occupation number over time.
\subsubsection{The a and b Modes}
As with the out-of-phase mode, the initial state must be pre-compensated before separation starts at $t=0$, but now with a suitable combination of single mode squeezing operations for $a$ and $b$ individually as well as a two-port interferometer. The initial and final normal mode frequencies in (\ref{Eq:decoupHam}) are 
\begin{align}
	\omega_a(0) &= \frac{\omega_0}{\sqrt{10}} \sqrt{13 + 21 \frac{m_D}{m_H} + \sqrt{\frac{512 m_D}{m_H} + \left(13 - 21 \frac{m_D}{m_H}\right)^2}} \nonumber \\
 &\equiv \alpha_a \omega_0, \nonumber\\
	\omega_a(t_f) &= \omega_0, \nonumber\\
	\omega_b(0) &= \frac{\omega_0}{\sqrt{10}} \sqrt{13 + 21 \frac{m_D}{m_H} - \sqrt{\frac{512 m_D}{m_H} + \left(13 - 21 \frac{m_D}{m_H}\right)^2}} \nonumber\\
 &\equiv \alpha_b \omega_0,\nonumber\\
	\omega_b(t_f) &= \sqrt{\frac{m_D}{m_H}} \omega_0 \equiv \beta_b \omega_0.
\end{align}
When applied to modes in their ground states initially, the interferometer $\boldsymbol{B}(\boldsymbol{\theta}_1)$ in the decomposition (\ref{Eq:EulerDecomp}) of the pre-compensation operation has no effect. It is sufficient to squeeze the two modes followed by an interferometer,
\begin{align}
   \boldsymbol{M}^{(p)}_{ab} &= \boldsymbol{B}_{ab}(\boldsymbol{\theta}) \left[\boldsymbol{S}_a(r_a,\phi_a) \oplus \boldsymbol{S}_b(r_b, \phi_{b})\right].
\end{align}
This can be further reduced by making use of (\ref{eq:twoModeInter}) and the insertion of rotation operators, remembering that rotations leave the ground state invariant,
\begin{align}
    \boldsymbol{M}^{(p)}_{ab} &= \boldsymbol{B}_{BS}(\theta_{BS}, \phi_{BS}) \cdot \boldsymbol{R}_{ab}(\theta_a, \theta_b) \left[\boldsymbol{S}_a(r_a,\phi_a) \oplus \boldsymbol{S}_b(r_b, \phi_{b})\right] \boldsymbol{R}_{ab}(-\theta_a, -\theta_b) \cdot \boldsymbol{R}_{ab}(\theta_a, \theta_b) \\
    &= \boldsymbol{B}_{BS}(\theta_{BS}, \phi_{BS}) \left[\boldsymbol{S}_a(r_a,\phi_a') \oplus \boldsymbol{S}_b(r_b, \phi_{b}')\right] \boldsymbol{R}_{ab}(\theta_a, \theta_b) \\
    &\to \boldsymbol{B}_{BS}(\theta_{BS}, \phi_{BS}) \left[\boldsymbol{S}_a(r_a,\phi_a') \oplus \boldsymbol{S}_b(r_b, \phi_{b}')\right]
\end{align}
The covariance matrix of the initial state is
\begin{align}
    \boldsymbol{V}^{(i)}_{ab} &= 
    \begin{pmatrix}
        \frac{\alpha_{a} \omega_0}{2} & 0 & 0 & 0 \\
        0 & \frac{\alpha_{b} \omega_0}{2} & 0 & 0 \\
        0 & 0 & \frac{1}{2\alpha_{a} \omega_0} & 0 \\
        0 & 0 & 0 & \frac{1}{2\alpha_{b} \omega_0}
    \end{pmatrix},
\end{align}
and that of the final state is
\begin{align}
    \boldsymbol{V}^{(f)}_{ab} &= 
    \begin{pmatrix}
        \frac{\omega_0}{2} & 0 & 0 & 0 \\
        0 & \frac{\beta_b \omega_0}{2} & 0 & 0 \\
        0 & 0 & \frac{1}{2 \omega_0} & 0 \\
        0 & 0 & 0 & \frac{1}{2 \beta_b \omega_0}
    \end{pmatrix}.
\end{align}
For the the $a$ and $b$ modes, $\boldsymbol{h}(t)$ in (\ref{Eq:generalHam}) takes the form
\begin{align}
    h_{ab}(t) &= \begin{pmatrix}
        1 & 0 & 0 & -\dot{\theta}(t) \\
        0 & 1 & \dot{\theta}(t) & 0 \\
        0 & \dot{\theta}(t) & \omega_a^2(t) & 0 \\
        -\dot{\theta}(t) & 0 & 0 & \omega_b^2(t)
    \end{pmatrix}.
\end{align}
The equation of motion for $\boldsymbol{M}_{ab}$ is
\begin{align}
    \dot{\boldsymbol{M}}_{ab}(t) &= - \boldsymbol{C} \cdot \boldsymbol{h}_{ab}(t) \cdot\boldsymbol{M}_{ab}(t) \\
    \boldsymbol{M}_{ab}(0) &=\boldsymbol{I}.
\end{align}
and can be solved numerically with $ \boldsymbol{M}_{ab}(t_f)\equiv  \boldsymbol{M}^{(f)}_{ab}$ and inverted to yield
\begin{align}
\boldsymbol{V}^{(p)}_{ab} \equiv (\boldsymbol{M}^{(f)}_{ab})^{-1} \cdot\boldsymbol{V}^{(f)}_{ab} \cdot[(\boldsymbol{M}^{(f)}_{ab})^{-1}]^T,
\end{align}
which can be compared to 
\begin{align}
\boldsymbol{V}^{(p)}_{ab} = \boldsymbol{M}^{(p)}_{ab} \cdot\boldsymbol{V}^{(i)}_{ab} \cdot[\boldsymbol{M}^{(p)}_{ab}]^T,
\end{align}
to find the correct parameters for pre-compensation numerically.\\
\\
The squeezing parameters required for GGS on the $a$ and $b$ modes in this BMB crystal are $r_a \approx 1.938$ and $r_b \approx 1.483$, respectively, with phases $\phi_{a} \approx -1.846$ and $\phi_{b} \approx -2.902$.  The required beam-splitting parameters are $\theta_B \approx 1.714$, and $\phi_B \approx -1.470$. In analogy to the out-of-phase mode, the occupation numbers for both modes can be written as
\begin{align}
    \expval{n_{a}}_t &= \frac{\frac12 V^{ab}_{11}(t) + \frac12 \omega_0^2 V^{ab}_{33}(t)}{\omega_0} - \frac12, \\
    \expval{n_{b}}_t &= \frac{\frac12 V^{ab}_{22}(t) + \frac12 (\beta_{b} \omega_0)^2 V^{ab}_{44}(t)}{\beta_{b} \omega_0} - \frac12,
\end{align}
and are shown in Fig.~\ref{fig:timeEvolvePlots}(c).
%
\subsection{Generalization to Larger Crystals} \label{sec:LargeCrystals}
The generalization of this procedure to larger crystals is not difficult but rather tedious.  For example, a crystal such as DHHD will have four modes that decouple into two groups of coupled oscillators. Each group will have its own rotation angle $\theta_{1,2}$ and appear in the same way as the rotation appears in (\ref{Eq:decoupHam}).  In this way, state preparation can be done on the two groups individually through two groups of single mode squeezers and beam-splitters.

The generalization to crystals such as D...DHD...D with $N$ ions, for odd $N$, is again straightforward, but requires a slightly more complicated decoupling procedure where we end up with two groups.  The first group contains $X=(N+1)/2$ coupled modes described by in-phase modes with participation from the $N-1$ data qubits as well as the helper ion.  The second group contains $Y=(N-1)/2$ coupled out-of-phase modes for the $N-1$ data qubits in which the helper ion does not participate.  The state preparation for such a crystal will require $X$ single mode squeezers and an $X$-port interferometer on the first mode group as well as $Y$ single mode squeezers and a $Y$-port interferometer on the second mode group.
\subsection{Time-scales for Pre-compensation}
Operations to pre-compensate the motional modes for GGS require a finite amount of time.  Resonant squeezing of a motional mode at frequency $\omega_k$ can be accomplished by modulating the potential curvature at $2 \omega_k$. To our knowledge, the strongest experimentally demonstrated squeezing rate for a single motional mode is $r=t/(3.2~\mu\text{s})$ \cite{Burd_2019}. This rate can potentially be increased, but to keep the effects on spectator modes negligible, the duration of each squeezing operation must be substantially above $1 /(2 \min[\Delta\omega_{kl}])$ where $\Delta\omega_{kl}$ is the frequency difference between modes $k$ and $l \neq k$. For mode frequencies between 1 MHz $\leq \omega_k \leq$ 3 MHz  we estimate a total duration not shorter than approximately 6 $\mu$s for all three squeezing operations. A balanced beam-splitter between two axial modes in a three ion BMB crystal was implemented in 17 $\mu$s in \cite{Hou2022qee} by driving a suitable coupling potential at $\Delta \omega_{kl}$. Similar considerations about the spectator mode as for resonant squeezing apply, and we can estimate a duration no shorter than 2 $\mu$s for the beam-splitter operation. Under these assumptions, the total duration of pre-compensation and separation is on the order of 10 $\mu$s, substantially larger than the 2 $\mu$s required for the three ions to separate by 80 $\mu$m as shown in Fig.~\ref{fig:timeEvolvePlots}.
\section{Three Ion GGS through on-the-fly compensation} \label{sec:fastThreeIon}
Since the pre-compensation removes effects from diabatic changes of the external potential, it seems reasonable that a more elaborate choice of dynamical external potential than just ramping down and back up can separate the ions {\it and} have all three modes start and end in their ground states in a duration that is on the same time-scale as the Coulomb expansion. This ``on-the-fly compensation'' will be explored next.\\
\\
\begin{figure}[h!]
\centering
\includegraphics[width=4.5in, height=6.25in]{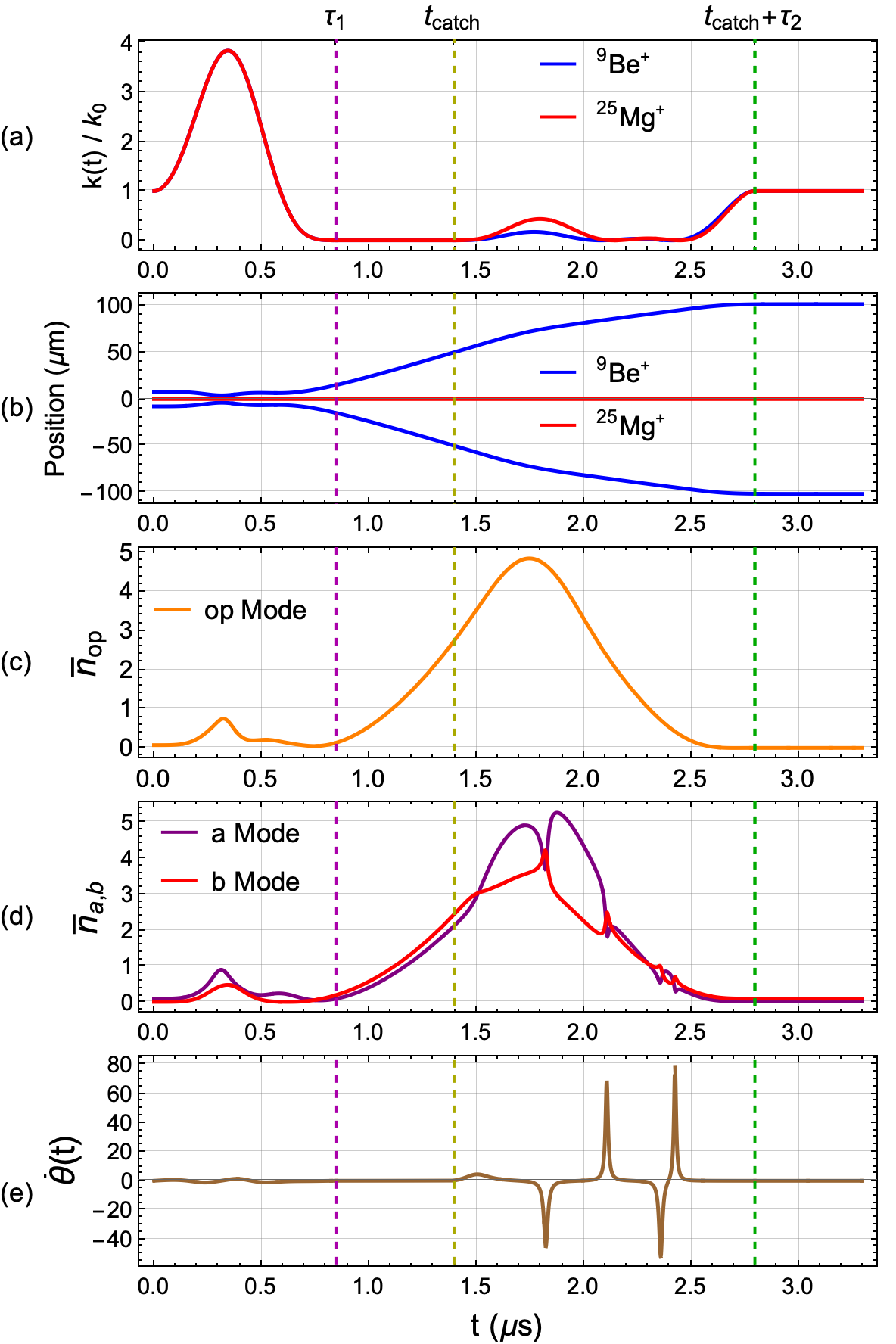}
\caption{Two numerical solutions. (a) Axial mode well curvature $k(t)/k_0$ for a single ion over time.  Red corresponds to the curvature a trapped single $^{25}$Mg$^+$ ion experiences, and blue corresponds to the curvature the $^9$Be$^+$ ions experience.  (b) Classical positions of the M ion (red) and the two B ions (blue). Initially the ions are pushed together and then rapidly move apart, while the trapping potential is modulated up and then ramps down to reach a minimal value (leftmost dashed vertical line in (a)-(e)). After the B ions reach $\approx \pm 50~\mu$m distance from the M position, individual potentials are turned on that modulate the curvature around the M and the B ions while  the B ions are simultaneously decelerated (center dashed line). When all ions come to rest, their occupation numbers are near zero and the potentials are no longer changing (rightmost dashed line). (c) Occupation number of the $op$ mode during the time evolution. (d) Occupation numbers of the $a$ and $b$ modes during time evolution. (e) Plot of $\dot{\theta}(t)$ as defined in (\ref{Eq:decoupHam}) and (\ref{eq:thetaDef}).  The mode interchanging in (d) is caused by the cusp-like behavior in $\dot{\theta}$.}\label{fig:timeEvolveFast}
\end{figure}
Our starting point is the partially decoupled Hamiltonian (\ref{Eq:decoupHam}), assuming a BMB crystal. Separation will be driven by lowering the external potential to zero, similar to the separation in \ref{sec:TransProt}.  In contrast to what is described in \ref{sec:TransProt}, the curvature of the trapping potential proportional to $k_D(t)$ is increased starting at $t=0$ to squeeze and couple the modes while the ions experience substantial Coulomb interaction. Additionally, this modulation decreases the distance between the ions and effectively ``tensions the springs'' before separation. The curvature is ramped up and down quickly over $\tau_{1} = 0.85~\mu$s to near zero confinement such that the ions fly apart driven by their mutual Coulomb repulsion.   As the trap is of small but nonzero strength with frequency $\omega_{op} = \omega_0/30$, the B ions reach a distance of $50 ~\mu$m from their starting equilibrium position at $t_{\text{catch}} \sim 1.4~\mu$s, far enough apart to create individual potential minima close to the position of all three ions. The individual potential curvatures are increased and further modulated, and the positions of the B minima move further apart to near $\pm 80~\mu$m ion distance while also providing a decelerating force that is proportional to the velocity of the B ions analogous to (\ref{Eq:friction_term}) to bring them to rest over $\tau_{2} = 1.4~\mu$s. The modulation of the curvature transforms the final modes back to near their ground states at $t = t_f = t_{\text{catch}} + \tau_2 \approx 2.8 ~\mu\text{s}$ after which the external potential is held constant. All put together, the time dependence of the well frequency that the B and M ions experience is
\begin{align}
\omega_{B,M}(t) &=
    \begin{cases}
        \omega_0 & t < 0\\
        \omega_{\text{down}}(t) & 0 \le t < \tau_{1} \\
        \omega_{op} = \omega_0/30 & \tau_{1} \le t < t_{\text{catch}} \\
        \omega_{\text{\text{catchB},\text{M}}}(t) & t_{\text{catch}} \le t < t_{\text{catch}} + \tau_2 \\
        \omega_0 & t_{\text{catch}} + \tau_2 \le t =t_f
    \end{cases}
\end{align}
where $\omega^2_{B}(t) = k_B(t)/m_{B}$ and $\frac{m_B}{m_M}\omega^2_M(t) = k_M(t)/m_M$. The scaling of $\omega^2_M(t)$ by the ratio of masses is so that both $\omega_M(0) = \omega_0$ and $k_M(0) = k_B(0) \equiv k_0$ hold at $t=0$.  We represent the time dependent parts of the  profile, $\omega_m(t)$ ($m \in \{\text{down}, \text{catchB}, \text{catchM}\}$), by a truncated Fourier series, which can provide a description of arbitrary time dependence with relatively few parameters (compared to other approaches such as splines) 
\begin{align} \label{eq:FourierSeries}
    \frac{\omega_m(t)}{\omega_0} &= a_0 + \sum_{\ell=1}^4\left(a_\ell \cos\frac{\pi \ell t}{2 \tau_{m}} + b_\ell \sin\frac{\pi \ell t}{2 \tau_{m}}\right)
\end{align}
with $a_\ell$ and $b_\ell$ being the Fourier components, $\tau_{m}$ the amount of time the potential is modulated for, and $\omega_0$ the initial well frequency.  For the first modulation, $\omega_{\text{down}}$, several of the Fourier components are constrained by the following boundary conditions,
\begin{align}
    \omega_{\text{down}}(0) &= \omega_0,
    \hspace{.1\linewidth}
    \omega'_{\text{down}}(0) = 0\\
    \omega_{\text{down}}(\tau_{1}) &= \omega_{op},
    \hspace{.1\linewidth}
    \omega_{\text{down}}'(\tau_{1}) = 0,
\end{align}
The minimum of the catching potential for the B moving in the positive direction moves along  $c_c(t) = c_B(t) - \eta \dot{c}_B(t)$, where $c_b(t)$ is the classical position of this B. For the B ion moving in the negative direction, the catching potential position is moving equal and opposite.  Hence, the catching potentials implement a force to slow the ions that is proportional to their classical velocity. The time dependence of the catching potential frequencies will be described by another truncated Fourier series (\ref{eq:FourierSeries}) with altered boundary conditions; initial frequency $\omega_{op}$ and final frequency $\omega_0$.\\
\\
If the unconstrained Fourier components of both the initial well modulation and catching potentials are chosen correctly, modes will be left in their approximate ground states at $t_f$. Fig. \ref{fig:timeEvolveFast} illustrates one example of this procedure. The Fourier components are,
\begin{align} \label{eq:Components}
    (a_0, a_1, a_2, a_3, a_4)_{\text{down}} &= (2.217, 0.3, -0.517, -0.5, -0.5) \nonumber\\
    (b_1, b_2, b_3, b_4)_{\text{down}} &= (-2.2, 0.1, 0.0, 0.5), \nonumber\\
    (a_0, a_1, a_2, a_3, a_4)_{\text{catchB}} &= (24.2, 172.7, -206.5, -0.5, 11.1) \nonumber\\
    (b_1, b_2, b_3, b_4)_{\text{catchB}} &= (-202.3,-0.1,9.5,43.5), \nonumber\\
    (a_0, a_1, a_2, a_3, a_4)_{\text{catchM}} &= (27.2, 229.5, -264.5, -0.5, 9.3) \nonumber\\
    (b_1, b_2, b_3, b4)_{\text{catchM}} &= (-260.5, -0.5, 10.5, 57.5).
\end{align}
The parameters for this modulated separation were found using a Nelder-Mead numerical optimization scheme such that the total occupation number ($\bar{n}_{op}$ and $\bar{n}_a + \bar{n}_b$ separately) was minimized.

A perfect implementation of the waveform described by (\ref{eq:Components}) will give final occupation numbers of $\bar{n}_{op} \sim 0.006$, $\bar{n}_a \sim 0.034$, and $\bar{n}_b \sim 0.11$.  As squeezed states in general can have long tails of occupation in the number basis, we also consider the number-basis populations $P_n = \abs{\braket{n}{\psi_f}}^2$ for the single $op$-mode case and $P_{n,m} = \abs{\braket{n_a m_b}{\psi_f}}^2$ in the coupled two mode ($ab$) case. We state the first few non-zero probabilities here for completeness:
\begin{center}
\begin{tabular}{||c||c|c|c|c|c||}
    \hline
     Mode & $P_0$ & $P_1$ & $P_2$ & $P_4$ & $P_6$\\
     \hline
     $\ket{\psi_{op}}$ & $0.997$ & $0$ & $0.00291$ & $1.28\times 10^{-5}$ & $6.22\times10^{-8}$\\
     \hline
     \hline
     $~$ & $P_{0,0}$ & $P_{1,1}$ & $P_{0,2}$ & $P_{2,0}$ & $P_{2,2}$ \\
     \hline
     $\ket{\psi_{ab}}$ & $0.937$ & $0.0219$ & $0.0326$ & 0.00297 & $0.00103$\\
     \hline
\end{tabular}
\end{center}
Here we see that states with odd total quantum number are disallowed as would be expected from squeezed states.  The squeezing parameters for each of the three modes can readily be determined from a Bloch-Messiah decomposition of the final states and are given by, $r_{op} \sim 0.0788$, $r_a \sim 0.0289$ and $r_b \sim 0.365$.

This example shows that modulation during separation is feasible and GGS separation can be executed on the same time scale that describes  the Coulomb expansion of a BMB crystal when the potential frequency is dropped from $\omega_0$ to zero (The B ions need approximately $1.3 ~\mu\text{s}$ to reach $\pm 80~\mu$m distance from the M ion in this case). The most time consuming portion in our example is catching the ions, and while we have not attempted this, it is likely that the entire separation could be sped up further with clever optimization, for example by increasing $\eta$ from the value of $0.4~\mu$s used in our example or by compressing the crystal even further before release.
\subsection{Occupation Number Response to Error in Fourier Components}
An important consideration for any transport or separation protocol is its robustness to error in its implementation. Here, we address this by introducing random deviations of the Fourier series coefficients.  These deviations are supposed to represent random uncontrolled noise, as opposed to static errors that can be compensated for by better calibration. We will introduce random deviations from the intended value in each component in (\ref{eq:Components}) sampled from a uniform distribution with maximum deviation equal to a fraction of the largest Fourier components in each waveform.  Results are presented for maximum fractional deviations of $10^{-5}$ and $5\times 10^{-5}$ of the largest Fourier components.  We Monte-Carlo sample from these distributions to produce average final state populations to characterize the robustness to error.

Recall that during the catching period of this algorithm, the catching potential minima followed the Be$^+$ ions through (\ref{Eq:friction_term}).  However, in a realistic scenario, this trajectory would be pre-computed based on an ideal implementation of the Fourier components in (\ref{eq:Components}) which are now disturbed by random error.  This random error causes our catching potentials to imperfectly slow down the ions, which will in turn oscillate in the final well.  The final state can be computed by performing a coherent displacement back into the lab frame through a displacement operator,
\begin{align}
    D_f &= \exp\bigg\{i\big[m_B(c_{B1}'(t_f) - w_{B1}'(t_f))x_{B1} + m_B(c_{B2}'(t_f) - w_{B2}'(t_f))x_{B2} \nonumber\\
    &- (c_{B1}(t_f) - w_{B1}(t_f))p_{B1} - (c_{B2}(t_f) - w_{B2}(t_f))p_{B2}\big]\bigg\}.
\end{align}
where $x_{Bi}$ are the \emph{original} unscaled operators.  For simplicity, we do not allow for any asymmetry between the two catching potentials during the slowdown period for the two Be$^+$
ions, i.e. $c_{B1} = -c_{B2}$.  This has the effect of canceling any classical motion effects at the end of transport on the $M$ ion's classical position as well as the $a$ and $b$ modes, which are essentially COM modes, while still providing a realistic assessment of the final state of the $op$ mode with noisy potentials.  Finally, if the classical trajectory is measured in units of meters and seconds, the displacement of the scaled $op$-mode operators is given by,
\begin{align}
    D_f\dagg x_{op} D_f &= x_{op} + \sqrt{\frac{2m_B}{\hbar}} (c_{B1}(t_f) - w_{B1}(t_f)) \\
    D_f\dagg p_{op} D_f &= p_{op} + \sqrt{\frac{2}{\hbar m_B}} m_B (c_{B1}'(t_f) - w_{B1}'(t_f)).
\end{align}
\begin{figure*}[ht!]
    \centering
    \includegraphics[width=5.5in, height=5.in]{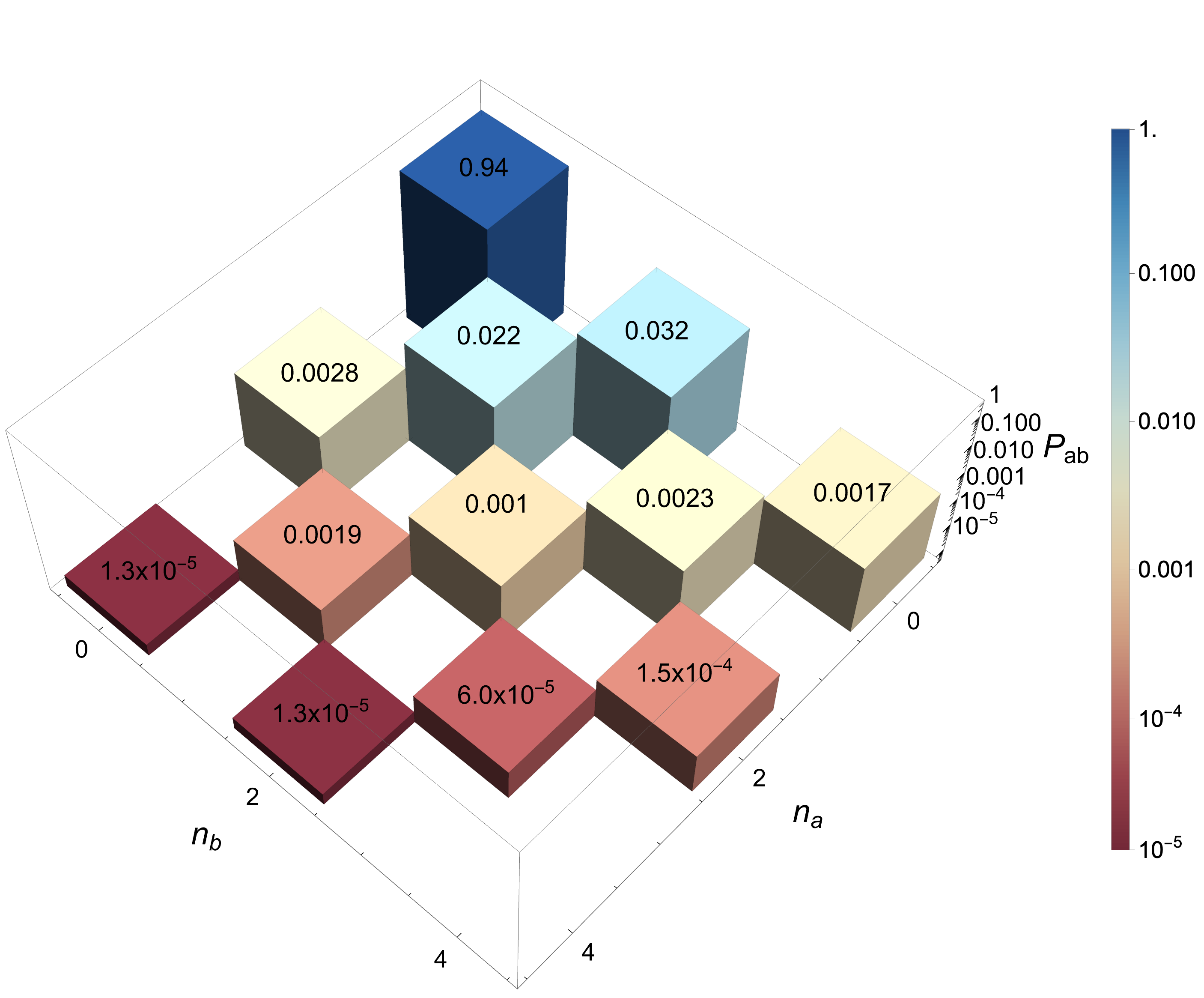}
    \caption{Monte-Carlo computed average final state occupations for the two coupled $a$ and $b$ modes.  A random offset is added to each Fourier component, with offsets uniformly distributed around zero with a maximum value equal to $10^{-5}$ of the largest Fourier component for each waveform.  The bar height, and value written on the bar, indicate the occupation of each joint motional state. The tallest blue bar in the upper corner is the probability for both modes to be in the ground state, i.e. $P_{00}$.  Moving down the chart, we find nonzero occupation probabilities for some higher excited states including $P_{22}$, $P_{24}$, and others.  For simplicity, we constrain the trapping potentials to be mirror symmetric in $x$ (including the random offset) during the catching phase, so the $a$ and $b$ modes (akin to COM modes) will not have occupation in odd number states, $P_{10} = P_{21} = \cdots = 0$.}
    \label{fig:colorHist}
\end{figure*}
\begin{table}[h!]
\centering
\begin{tabular}{||c||c|c|c|c||}
    \hline
      Error & $P_0$ & $P_1$ & $P_2$ & $P_3$\\
     \hline
     \hline
     $10^{-5}$ & $0.97$ & $0.025$ & $0.003$ & $0.0002$ \\
     \hline
     $5\times10^{-5}$ & $0.63$ & $0.12$ & $0.056$ & $0.033$ \\
     \hline
\end{tabular}
\caption{Table of Monte-Carlo computed average occupations of the $op$-mode when the maximum error is $10^{-5}$ or $5\times 10^{-5}$ of the largest Fourier components.} \label{table:ErrorOpMode}
\end{table}

\begin{table}[h!]
\centering
\begin{tabular}{||c||c|c|c|c|c||}
    \hline
     \backslashbox{$P_a$}{$P_b$} & $0$ & $1$ & $2$ & $3$ & $4$\\
     \hline
     \hline
     $0$ & $0.94$ & $0$ & $0.032$ & $0$ & $0.0017$\\
     \hline
     $1$ & $0$ & $0.022$ & $0$ & $0.0023$ & $0$\\
     \hline
     $2$ & $0.0028$ & $0$ & $0.0010$ & $0$ & $0.00015$\\
     \hline
     $3$ & $0$ & $0.00019$ & $0$ & $0.000060$ & $0$\\
     \hline
     $4$ & $0.000013$ & $0$ & $0.000013$ & $0$ & $\approx 0$\\
     \hline
\end{tabular}
\caption{Table of Monte-Carlo computed average transition probabilities of the final state for the $a$ and $b$ modes when the error is $10^{-5}$ of the largest Fourier components.  The equivalent table produced when the error is taken to be $5\times 10^{-5}$ is essentially indistinguishable from this. These data are plotted in Fig. \ref{fig:colorHist}.} \label{table:ErrorABMode1}
\end{table}
Table \ref{table:ErrorOpMode} shows the average final transition probability of the $op$ mode while Fig. \ref{fig:colorHist} and Table \ref{table:ErrorABMode1} show the average final transition probabilities of the $a$ and $b$ modes when the error is $1\times10^{-5}$ and $5\times 10^{-5}$ of the largest Fourier components.  Notably, there is non-negligible occupation in the $n=1$ state of the $op$ mode, which implies that classical motion cannot be neglected even for such a small amount of noise in the waveforms.  If the final classical motion were to be neglected, the final occupations of the $op$ mode would be more akin to the $a$ and $b$ modes where the only residual occupation is due to imperfect squeezing and mode-mixing.  The second row in Table \ref{table:ErrorOpMode} makes this even more apparent when the error is allowed to increase to $5\times10^{-5}$ and we see non-negligible occupation of the $n=3$ state.  Here we again note that no asymmetry has been introduced into the B ion catching potentials which would disturb the classical motion of the M ion.  This has been done for simplicity, so any real implementation of this protocol must take asymmetric perturbations into account as states like $P_{0,1}$ and $P_{1,0}$ will have non-zero and non-negligible occupation.  We also note that the same precision in realizing the external potentials is required in implementations with pre-compensation and simple ramps, such as the one discussed in section \ref{sec:TransProt} and \cite{Sutherland:2021ifb}.

This analysis highlights the exquisite control over the Fourier components of the waveform necessary to reliably achieve fast GGS. Slower implementations will improve the robustness, with one limit represented by the essentially adiabatic separation methods that have already been demonstrated. Initial experimental implementations will likely be only slightly faster than adiabatic and durations will drop as control improves. Small deviations from the ground state can be removed by re-cooling the DHD crystal on the H ion \cite{Hou_2024} but recooling must be rapid to not defeat the purpose of fast separation, and therefore the modes should have as low as possible occupations after separation.  
%
\section{Conclusions} \label{sec:Conclusion}
Rapid separation of a DHD three ion crystal leads to squeezing and mode-mixing that needs to be carefully managed to end in the quantum mechanical ground states of all three axial normal modes of the separated ions. To achieve GGS, the effects of separation can either be pre-compensated by suitable squeezing and mode-mixing operations before  separation, or more expediently be mitigated by precisely controlled modulation of the potential during separation. Importantly, we find that there is not a significant time penalty for the latter method when compared to Coulomb expansion after the external trapping potential is instantaneously set to zero. An example implementation uses an efficient Fourier representation of the trapping potential and ideally terminates with all three axial modes close to the ground state.  The time dependence of the external potential is smooth, but requires very precise control that will take effort to implement with realistic driving electronics. Uncontrolled stray potential fluctuations will also need to be carefully minimized. Utilizing the Coulomb repulsion to drive the ions apart rather than narrow separation electrodes that create a ``wedge'' between ions \cite{Home_2006} may help with reducing the complexity and electrode count of traps that are suitable for GGS separation. Separating D ions out of a DHD crystal trapped in a single well allows for the D ions to travel through a larger trap array on their own so they can be paired with other D ions in subsequent gate operations, while the H ions stay in place. This approach could mitigate issues from transporting groups of ions of different mass through junctions in a large ion trap array \cite{Burton_2022} and simplifies transport in general, since all transport primitives besides separation only need to be implemented for the D species. In this context, it is worth noting that recombination of two D ions that approach a stationary H from opposite directions is the time reversal of DHD separation.  Since the equations of motion are invariant under time reversal, ground state to ground state recombination can be accomplished by running the separation protocol backwards in time. Performing imperfect separation and recombination multiple times will lead to error accumulation from imprecision in the electrode control, higher order than quadratic terms in the potential energy and fluctuating stray potentials, but small amounts of excess energy can be removed by cooling the H ion when the DHD crystal is confined in a single external well \cite{Hou_2024}. Further refinement of GGS can reduce the duration of subsequent cooling to a minimum.\\
\\
Although we do not explicitly show how the procedure in \ref{sec:LargeCrystals} generalizes to crystals with more than three ions, we describe its general implementation and how the effects of separation can be decomposed into single mode squeezing and multi-port interferometers. In general, external potential modulations can accomplish diabatic squeezing and mode-mixing operations not just for ground states but could be used for implementing complex Gaussian operations on any initial motional state of groups of ions, with the potential for further generalization to all 3$N$ motional modes of $N$ ions. This could be of interest in realizing error correction codes and quantum logical operations on bosonic qubits realized in the motion of ion crystals \cite{GKP_original,Fluhmann_2019} and for realizing entangled states of ion internal degrees of freedom generated by coupling them to non-classical states of the motion with Jaynes-Cummings type interactions \cite{Wineland_1994}.

\begin{acknowledgments}
The authors would like to thank Tyler Sutherland for useful discussions.  The work of DL and DS was supported by the NIST Quantum Information Program.  The work of TG and SL was performed under the auspices of the U.S. Department of Energy by Lawrence Livermore National Laboratory under contract DE-AC52-07NA27344.
\end{acknowledgments}

\bibliography{references.bib}

\end{document}